\documentclass[doublecol]{epl2} 

\usepackage{xcolor}
\usepackage{ulem}

\def\gsim{\mathrel{\rlap{\lower4pt\hbox{\hskip1pt$\sim$}}
    \raise1pt\hbox{$>$}}}       

\def\be{\begin{equation}}
\def\bea{\begin{eqnarray}}
\def\ee{\end{equation}}
\def\eea{\end{eqnarray}}

\def\z{\vert 0 \rangle}
\def\oo{\vert 1 \rangle}
\def\o{\vert 1 \rangle}

\title{A Brief History of Hawking's Information Paradox}
\shorttitle{A Brief History of Hawking's Information Paradox} 

\author{Xavier Calmet \thanks{E-mail: \email{x.calmet@sussex.ac.uk}}\inst{1} \and Stephen~D.~H.~Hsu\thanks{E-mail: \email{hsusteve@gmail.com}} \inst{2}}
\shortauthor{X. Calmet and S.D.H. Hsu}

\institute{                    
  \inst{1} Department of Physics and Astronomy,\\
University of Sussex, Brighton, BN1 9QH, United Kingdom\\
  \inst{2} Department of Physics and Astronomy\\ Michigan State University, East Lansing, Michigan 48823, USA
}

\abstract{
In this invited review, we describe Hawking's information paradox and a recently proposed resolution of it. Explicit calculations demonstrate the existence of quantum hair on black holes, meaning that the quantum state of the external graviton field depends on the internal state of the black hole. Simple quantum mechanics then implies that Hawking radiation amplitudes depend on the internal state, resulting in a pure final radiation state that preserves unitarity and, importantly, violates a factorization assumption which is central to the original paradox. Black hole information is encoded in entangled macroscopic superposition states of the radiation.}

\begin{document}

\maketitle
\section{Introduction and Formulation of Paradox}


The thermal nature of his radiation led Stephen Hawking in 1976 to argue that black holes destroy quantum information \cite{Hawking1976}\footnote{ Given the length limit for a Perspective article, we will not be able to review all the work since Hawking's seminal paper: see e.g.\cite{tHooft:1995qox,Hawking:2005kf,Hawking:2014tga,Hawking:2015qqa,Giddings:1995gd,Bousso:2022ntt,Engelhardt:2022qts}. Rather, we will focus on the material that is directly relevant to recent advances involving quantum hair.}. More precisely, Hawking argued that black holes cause pure states to evolve into mixed states. Quantum information that falls into a black hole does not escape in the form of radiation. Rather, it vanishes completely from our universe, thereby violating unitarity in quantum mechanics. 

In 2009 Mathur \cite{Mathur1,Mathur2} gave an especially clear formulation of the paradox. His analysis tracks the entanglement entropy of Hawking radiation emitted on a nice slice. Nice slices are spacelike surfaces which intersect both the interior of the black hole and the emitted Hawking radiation. Mathur emphasizes how the Hawking radiation originates from vacuum fluctuations of entangled particle and antiparticle states. The vacuum state near the horizon of a large black hole (i.e., where curvature is negligible) must be approximately that of the ordinary vacuum. One of the entangled pair (the Hawking radiation) escapes to infinity, while the other particle falls into the black hole. When the black hole finally evaporates and disappears from our universe, the radiation modes are left in a mixed state with very large entanglement entropy. 

Let us call the modes outside the horizon  $b_1, b_2, ...$, and those inside the horizon  $e_1, e_2, ...$. The initial slice in the nice slice foliation contains only the matter state $|\psi\rangle_M$ (i.e., the black hole), and none of the $e_i, b_i$. The first step of evolution stretches the spacelike slice, so that the particle modes $e_1, b_1$ are now present on the new slice. The state of these modes has the schematic form
\begin{equation}
\frac{1}{\sqrt{2}} \Big ( \z_{e_1}\z_{b_1}+ \oo_{e_1}\oo_{b_1}\Big )~,
\label{three}
\end{equation}
where the numbers $0, 1$ give the occupation number of a particle mode. The entanglement of the state outside the horizon (given by the mode $b_1$) with the state inside  (given by the mode $e_1$) yields 
$S_{\rm entanglement}=\ln 2$.

In the next step of evolution the modes $b_1, e_1$  at the earlier step move apart, and in the region between them there appears another pair of modes $b_2, e_2$ in a state that has the same form as (\ref{three}). After $N$ steps we have the state
\begin{eqnarray}
\vert \Psi \rangle \approx ~ |\psi\rangle_M&\otimes& \frac{1}{\sqrt{2}} \Big( \z_{e_1}\z_{b_1}+ \oo_{e_1}\oo_{b_1}\Big)\cr
&\otimes& \frac{1}{\sqrt{2}} \Big( \z_{e_2} \z_{b_2} + \oo_{e_2} \oo_{b_2} \Big)\cr
&\dots&\cr
&\otimes& \frac{1}{\sqrt{2}} \Big( \z_{e_N} \z_{b_N} + \oo_{e_N} \oo_{b_N} \Big).~
\label{six}
\end{eqnarray}
The state $\Psi$ has the tensor product form given above because each pair creation (Hawking emission) is widely separated from the others. There is no connection to the matter state in the leading order Hawking process. It is the factorized form of $\Psi$ that leads to the paradox, once the black hole and $e_i$ modes are gone.
This formulation has been further developed in \cite{firewall1,firewall2}, though it is fair to say that all formulations of the paradox are based on assumptions related to the  factorization of states apparent in (\ref{six}).

The modes $ b_i$ are entangled with the  $e_i$ and the black hole with $S_{\rm entanglement}=N\ln 2$.
This entanglement grows by $\ln 2$ with each succeeding emission. If the black hole evaporates away completely, the $b_i$ quanta outside will be in an entangled state, but there will be nothing that they are entangled {\it with}. The initial pure state has evolved to a mixed state, described by a density matrix.

Mathur argues further that small corrections $\epsilon$ to Hawking evaporation cannot change this qualitative result: the entanglement entropy increases by $\approx \ln 2 - \epsilon$ with each emitted quantum. We note, however, that this entire analysis has been specific to a given spacetime geometry which defines (on each nice slice) the location of the black hole horizon, the nature of the $e_i$ modes, etc. Nowhere in this analysis has the possibility been considered that there might be {\it other} background geometries involved: e.g., with black hole center of mass at a different location, or with different recoil velocity, or equivalently with a macroscopically different pattern of radiation $\{ b_1, b_2,..., b_N \}$ emitted. The black hole trajectory is determined by recoil from the emitted radiation, so different patterns correspond to different trajectories. Mathur's analysis does not consider whether small $\epsilon$ entanglements {\it between} macroscopically distinct radiation states can unitarize the evolution of $\Psi$. This will be examined further below.

\section{AdS/CFT and Holography}

The holographic principle -- i.e., that bulk information in models of gravity in $d$-dimensions might be available on the $(d-1)$ dimensional boundary of spacetime -- is due to 't Hooft (1993) \cite{tHooft:1993dmi} and Susskind (1995) \cite{Susskind:1994vu}. Holography was given an explicit realization in the AdS/CFT correspondence of Maldacena (1998) \cite{Maldacena:1997re}, which suggests that black hole evaporation can be unitary. That is, formation and evaporation of a small black hole in the bulk is dual to the (presumably unitary) evolution of some gauge configuration on the boundary. Indeed, as conventionally interpreted, the AdS/CFT correspondence provides a wide range of examples of unitary quantum dynamics of black holes. Recently there has been considerable progress in directly computing the entanglement entropy of evaporation using AdS methods\footnote{Under Schr\"odinger evolution the quantum state describing a small black hole in AdS will become a macroscopic superposition state with the hole on different evaporation trajectories. Presumably the gauge dual of this bulk state evolves similarly. Neither outcome is surprising, given the results in \cite{Buniy:2020dux}. It would be interesting to explore whether these observations are reflected in bulk-boundary duality \cite{Almheiri:2020cfm}.}, and these results suggest that the process is unitary \cite{Almheiri:2020cfm}. 

However, the precise mechanism by which black hole information is encoded in outgoing radiation quanta has never been identified. To fully resolve the paradox requires an understanding of how the bulk gravitational dynamics is modified so that it becomes unitary. Various scenarios have been advocated over time, including exotic scenarios with gross violations of locality near the horizon, or stable black hole remnants. It is unclear whether exotic new physics is required to resolve the paradox -- our analysis below suggests that it is not.

Recently, the authors of \cite{R1,R2,R3,Raju:2020smc} have argued that bulk information is recoverable near the boundary of spacetime via local measurements. They do not assume a specific short distance completion of quantum gravity, but make some plausible assumptions, such as that the spectrum of energy eigenvalues is bounded below. If correct, these results provide a concrete realization of holography that the authors refer to as holography of information. 

Holography of information implies that the internal quantum state of a black hole must be encoded in the asymptotic quantum state of its graviton field, since otherwise the information would not be recoverable at the boundary. 
However, the details of how this works do not appear in \cite{R1,R2,R3,Raju:2020smc}.  
The notion of quantum hair that we introduce below enables us to address this question.

The idea that information cannot be localized in quantum gravity, and hence is recoverable at the boundary of spacetime, can be regarded as a consequence of the Gauss Law constraints on quantization, combined with the fact that mass (i.e., gravitational charge) cannot be screened\footnote{Absence of gravitational screening in the low-energy effective theory of gravity is commonly assumed, although mathematical studies of exotic (perhaps unphysical) spacetimes may suggest otherwise \cite{Carlotto}. We show explicitly \cite{Calmet:2021stu,Calmet:2021cip} that information about the black hole internal state is available in the quantum state of its gravity field (quantum hair).}. Note this assumes the absence of negative mass states -- i.e., a lower bound on the energy spectrum in quantum gravity. An alternative argument can be given using diffeomorphism invariance, which implies the absence of local observables in gravitational theories, see e.g.  \cite{Giddings:2005id,Kiefer:2014sfr} and references therein.

While both holography of information and AdS/CFT suggest that the Hawking paradox is somehow resolved in favor of unitarity, neither yield a specific description of the physical process by which black hole information is encoded in Hawking radiation which originates outside the horizon. This question has remained a mystery over the 20+ years that have elapsed since Maldacena gave the first examples of bulk-boundary duality.
As evidence supporting this conclusion, see these recent reviews \cite{Bena:2022ldq,Martinec:2022lsb}.

\section{Quantum Hair and Unitary Evaporation}

We shall now describe the solution to Hawking's paradox proposed in \cite{Calmet:2021cip}.
In \cite{Calmet:2021stu}, the analysis of the state of the graviton field produced by a compact matter source (e.g., a black hole) revealed the following: 

1. The asymptotic graviton state of an energy eigenstate source is determined at leading order by the energy eigenvalue, and can be expressed explicitly as a coherent state which depends on this eigenvalue. Insofar as there are no accidental energy degeneracies there is a one to one map between graviton states and matter source states\footnote{It is a folk theorem in many-body physics that typical energy level splittings are $\sim \exp(-S)$ ($S$ is the entropy of the system) and there are no exact degeneracies without exact symmetries. Even if distinct but exactly degenerate energy eigenstates $\psi_1$ and $\psi_2$ exist, it still seems likely that they are distinguishable via their effects on the graviton quantum state since by assumption the functions $\psi_1 (x)$ and $\psi_2 (x)$ are not identical. These effects are subleading relative to the dependence on the energy eigenvalue.}. A semiclassical matter source produces an entangled graviton state.

2. Quantum gravitational fluctuations (i.e., graviton loops) produce corrections to the long range potential (e.g., $\sim r^{-5}$) whose coefficients depend on the internal state of the source. This provides an explicit example of how the graviton quantum state (corresponding to the semiclassical potential) encodes information about the internal state of a black hole. Note the calculation is insensitive to short distance effects in quantum gravity (i.e., the short distance completion of the model).

Using these properties, we can write the quantum state of the exterior metric (equivalently, the quantum state of the exterior geometry) as 
\begin{equation}
 \Psi_i = \sum_n c_n \Psi_g (E_n) = \sum_n c_n \, \vert \, g(E_n) \, \rangle  ~.
\end{equation}
A semiclassical state has support concentrated in some range of energies $E$, where the magnitudes of $c_n$ are largest. For simplicity, when representing the exterior metric state $g(E)$ we only write the energy explicitly and suppress the other quantum numbers. 

Assume for convenience that the black hole emits one quantum at a time (e.g., at fixed intervals), culminating in a final state of $N$ radiation quanta:
$$ \vert ~ r_1 ~ r_2 ~ r_3 ~ \cdots ~ r_N ~ \rangle ~.$$
The quantum numbers of the $i$-th emitted radiation particle include the energy $\Delta_i$, momentum $p_i$, spin $s_i$, charge $q_i$, etc. The symbol $r_i$ is used to represent all of these values: $$r_i \sim \{ \Delta_i, p_i, s_i, q_i, \ldots \} ~.$$
A final radiation state is specified by the values of $\{r_1, r_2, \ldots , r_N \}$.

Let the amplitude for emission of quantum $r_i$ from exterior metric state $\Psi_g (E)$ be $\alpha (E,r_i)$. This amplitude must approximate the semiclassical Hawking amplitude for a black hole of mass $E$. In the leading approximation the amplitudes are those of thermal emission, but at subleading order (i.e., $\sim S^{-k}$ for perturbative corrections such as those calculated in \cite{Calmet:2021stu}, or $\exp( -S)$ for nonperturbative effects, where $S$ is the black hole entropy) additional dependence on $(E, r_i)$ will emerge. The fact that these corrections can depend on the internal state of the hole is a consequence of quantum hair. It has been shown that even corrections as small as $\exp(-S)$ can purify a maximally mixed Hawking state (i.e., can perturb the radiation density matrix $\rho$ so that ${\rm tr} \, \rho^2 = 1$), because the dimensionality ($\sim \exp S$) of the Hilbert space is so large \cite{Papadodimas:2012aq,Hsu:2013cw,Hsu:2013fra,Bao:2017who}.

When the black hole emits the first radiation quantum $r_1$ it evolves into the exterior state given on the right below:
\begin{equation}
    \Psi_i ~\rightarrow~ \sum_n \sum_{r_1} c_n ~ \alpha (E_n, r_1) ~ \vert \, g(E_n - \Delta_1), r_1 \rangle 
    \label{evolution1} ~~.
\end{equation}
In this notation $g$ refers to the exterior geometry and $r_1$ to the radiation. The next emission leads to
\begin{eqnarray}
    \sum_n \sum_{r_1,r_2} c_n ~ \alpha (E_n, r_1) ~ \alpha (E_n - \Delta_1, r_2) \times  \nonumber \\ 
    \vert \, g(E_n - \Delta_1 - \Delta_2), r_1, r_2 \rangle 
    \label{evolution2} ~~,
\end{eqnarray}
and the final radiation state is
\begin{eqnarray}
\label{evolution3}
\sum_n \sum_{r_1,r_2,\ldots,r_N} c_n ~\alpha(E_n, r_1) ~\alpha(E_n - \Delta_1, r_2) ~ \times  \nonumber \\ 
\alpha(E_n - \Delta_1 - \Delta_2, r_3 ) ~
\cdots ~ \vert \, r_1 ~ r_2 ~ \cdots ~ r_N \rangle ~. 
\end{eqnarray}
In the final state we omit reference to the geometry $g$ as the black hole no longer exists: there is no horizon and the spacetime is approximately flat. 

As is the case with the decay of any macroscopic object into individual quanta, several remarks apply.
A semiclassical black hole state has support concentrated in a small range of energy eigenvalues $E_n$. Realistic radiation quanta, which are localized in space and time, are better represented as wave packet states than as plane wave states, and hence have a small spread as well in their energies $\Delta_i$. The energy conservation requirement: $E_n \approx \sum_i \, \Delta_i$ is not exact.

The final expression (\ref{evolution3}) is the same one we would obtain from a burning lump of coal if we interpret the amplitudes $\alpha (E, r)$ with $r$ (again) the state of the radiated particle and $E$ the state of the coal as it burns. 

Quantum hair allows the internal state of the black hole, reflected in the coefficients $c_n$, to affect the Hawking radiation. The result is manifestly unitary, and the final state in (\ref{evolution3}) is manifestly a pure state:


$\bullet$ For each distinct initial state given by the $\{ c_n \}$ there is a different final radiation state.

$\bullet$ The time-reversed evolution of a final radiation state results in a specific initial state.


Note the final radiation state in (\ref{evolution3}) is a macroscopic superposition state. Configurations described by $\{r_1, r_2, \ldots , r_N \}$ will exhibit different regions with higher or lower densities of energy, charge, etc. Time-reversed evolution will only produce the original semiclassical black hole state if these radiation states converge and interfere with the exact phase relations given in (\ref{evolution3}). 

We emphasize that the quantum hair provides a mechanism by which the  amplitudes $\alpha (E, r)$ can depend on the internal state. These results provide, for the first time, a physical picture of how the information is encoded in the Hawking radiation. 

Specifically, the amplitudes $\alpha (E, r)$ can be explicitly calculated using the quantum corrected metric derived in \cite{Calmet:2021stu,Calmet:2019eof} using effective field theory methods. We expect a correction to Hawking's thermal radiation spectrum which is dependent on the composition of the original star via the corrections encoded in the $G_N^2 r^{-5}$ quantum gravitational corrections to the potential. This explicitly shows how Hawking radiation retains a memory of the original matter configuration that collapsed to a black hole.

\section{Mathur's Theorem revisited}

Recall that in Mathur's formulation of the paradox each vacuum fluctuation contains an entangled pair $(e_k, b_k)$, with the $b_k$ modes escaping to infinity as Hawking radiation, and the $e_k$ modes falling into the black hole. These modes are defined with respect to a specific background geometry which has to be consistent with the overall coarse grained pattern of emitted radiation.

Two radiation patterns $\{ b_1, b_2, ... , b_N \}$ and $\{ b'_1, b'_2, ... , b'_N \}$ which are sufficiently different correspond to distinct semiclassical geometries and hence distinct nice slices. The existence of distinct geometries becomes apparent as soon as we realize that the black hole recoil trajectory during evaporation is determined by the radiation pattern. The set of possible recoil trajectories is a collection of random walks, and the eventual spread in location of the hole is macroscopic, of order $\sim M^2$.

This formulation of the paradox does not address the possibility of entanglement {\it between} different geometries (i.e., between sufficiently different patterns of radiation as described above). Furthermore, the nature of the infalling $e$ modes is different on each geometry as they are excitations with respect to a specific nice slice.

The $b$ modes are the $r$ modes in Eq. (\ref{evolution3}). In our description the entanglement between $b$ and $e$ modes is reflected in the entanglement between $r$ modes and the exterior metric state $g(E)$, which is itself determined by the internal black hole state. 

A negative energy $e$ mode reduces the total energy of the internal state: $E \rightarrow E - \Delta$, and this is then reflected in the exterior metric state $g(E - \Delta)$. Other quantum numbers such as spin or charge which characterize the interior state are similarly modified by the infalling $e$ modes. 

The analog of the entangled state $\Big ( \z_{e_i}\z_{b_i}+ \o_{e_i}\o_{b_i}\Big )$ is simply
\begin{equation}
\vert \, g(E - \Delta), r \, \rangle ~+~ \vert \, g(E - \Delta'), r' \, \rangle~~.
\end{equation}
Upon inspection, it is easy to see that our final radiation state (\ref{evolution3}) contains many ``copies'' of the Mathur state, corresponding to different background geometries, which are themselves determined by the coarse grained stress tensor of a specific radiation pattern. However, our expression is manifestly unitary and exhibits entanglements across different background geometries, or equivalently between radiation patterns that differ macroscopically. These entanglements do not appear anywhere in Mathur's formulation and are responsible for unitarization of the evaporation process.

\section{Pure vs Mixed States and Concentation of Measure}

Let $Y$ be the Hilbert space describing all possible radiation patterns $\{ r_1, r_2, \cdots r_N \}$. The subset of radiation patterns $X$ which is consistent with a specific semiclassical geometry (i.e., nice slice) is a very small subset $X$ of the final state Hilbert space $Y$. 

Concentration of measure (Levy's lemma) \cite{Buniy:2020dux} implies that for almost all pure states $\Psi$ in a large Hilbert space $Y$, the density matrix 
\begin{equation}
\rho (X) =  {\rm tr} ~\Psi^\dagger \, \Psi     
\end{equation}
describing a small subset $X$ (tracing over the complement of $X$ in $Y$) is nearly maximally mixed -- i.e., like the radiation found in Hawking's original calculation. That is, a pure final state in $Y$ resulting from black hole evaporation might appear to be nearly maximally mixed to an observer (i.e., measuring devices) restricted to a specific, fixed background geometry. 

The fact that pure states in large Hilbert spaces are very difficult to distinguish from mixed states is central to the idea of decoherence: exponentially sensitive measurements are required to determine whether pure states evolve into mixed states (as in the case of wavefunction collapse) or merely appear to (decoherence) \cite{Hsu:2009ve}.

In almost all discussions of the black hole information paradox, entanglement between different $X$ and $X'$ (equivalently, between different branches of the wavefunction $\Psi$) has been neglected, although even exponentially small entanglement between these branches may be sufficient to unitarize the result \cite{Papadodimas:2012aq,Hsu:2013cw,Hsu:2013fra}.

The final state in Eq. (\ref{evolution3})
\begin{eqnarray}
\sum_n \sum_{r_1,r_2,\ldots,r_N}  c_n ~\alpha(E_n, r_1) ~\alpha(E_n - \Delta_1, r_2) ~   \nonumber \\
\alpha(E_n - \Delta_1 - \Delta_2, r_3 ) ~ 
 \cdots ~  \vert \, r_1 ~ r_2 ~ \cdots ~ r_N \rangle
 \nonumber
\end{eqnarray}
exhibits such $X$-$X'$ entanglements: it does not factorize across semiclassical spacetimes.

As we noted previously, even corrections as small as $\exp(-S)$ can purify a maximally mixed Hawking state (i.e., can perturb a maximally mixed radiation density matrix $\rho$ so that ${\rm tr} \, \rho^2 = 1$), because the dimensionality of the Hilbert space $Y$ ($\sim \exp S$) is so large.

\section{Final Remarks} 

Consider a burning lump of coal instead of an evaporating black hole. Following a similar analysis we obtain an expression for the final radiation state like Eq. (\ref{evolution3}) except that $\alpha(E , r)$ is the amplitude for the coal state with energy eigenvalue $E$ to emit radiation state $r$. Due to fluctuations in the (coarse grained) temperature and pressure of the emitted radiation, the initial coal state also evolves into a macroscopic superposition of radiation states. This is not surprising: it has been shown using von Neumann's 1929 Quantum Ergodic Theorem \cite{vNeumann} that under very weak assumptions generic initial states will evolve into macroscopic superposition states \cite{Buniy:2020dux}.

Because our expressions (\ref{evolution1})-(\ref{evolution3}) are manifestly unitary, and the final radiation state is pure, the Page curve \cite{Page:1993wv,Page:2013dx} of entanglement entropy as a function of evaporation time returns to zero. The entanglement entropy of the radiation state increases from zero, reaches a maximum while the evaporation is in progress, and returns to zero when the black hole has vanished and only the final pure state (\ref{evolution3}) remains.

The radiation amplitudes computed by Hawking, which describe thermal radiation emitted from a black hole at temperature $T$, already describe a broad distribution of possible radiation types, spins, and momenta emitted at each stage. Thus, even in the semiclassical approximation there are many distinct patterns of radiation in (\ref{evolution3}). The set of possible final states is already complex even at leading order, resulting in very different coarse grained patterns of energy-momentum density. Small corrections to the amplitudes $\alpha (E,r)$ due to quantum hair do not qualitatively change this situation, but they are necessary to unitarize the evaporation and they determine the precise relations between components of the entangled state.

Importantly, no special assumptions about the amplitudes $\alpha (E,r)$ need to be made to determine that the factorized form of the state (\ref{six}) does not hold. Factorization is assumed in essentially every formulation of the information paradox, but in reality is violated because the external graviton state depends on the internal black hole state. 

Known quantum gravitational effects leading to quantum hair are extremely small and thus difficult to probe experimentally or detect via observations. We cannot prove that our solution to the information paradox is unique. However, the consequences of quantum hair lead, {\it without any speculative theoretical assumptions}, to plausible unitary evaporation of black holes. The properties of quantum hair and the evaporation amplitude (\ref{evolution3}) can be deduced using only long wavelength properties of quantum gravity -- they do not rely on assumptions about Planck scale physics or a specific short distance completion. Therefore, Occam's razor favors quantum hair.

\acknowledgments
The work of X.C. is supported in part  by the Science and Technology Facilities Council (grants numbers ST/T00102X/1 and ST/T006048/1).

{\it Data Availability Statement:}
This manuscript has no associated data. Data sharing not applicable to this article as no datasets were generated or analysed during the current study.

\end{document}